

\message{   *** Please respond `b' to the query! ***     }

\input harvmac.tex
\input tables.tex
%
\global\newcount\mthno \global\mthno=1
\global\newcount\mexno \global\mexno=1
\global\newcount\mquno \global\mquno=1
\def\newsec#1{\global\advance\secno by1\message{(\the\secno. #1)}
\global\subsecno=0\xdef\secsym{\the\secno.}\global\meqno=1\global\mthno=1
\global\mexno=1\global\mquno=1
\bigbreak\medskip\noindent{\bf\the\secno. #1}\writetoca{{\secsym} {#1}}
\par\nobreak\medskip\nobreak}
\xdef\secsym{}
\global\newcount\subsecno \global\subsecno=0
\def\subsec#1{\global\advance\subsecno by1\message{(\secsym\the\subsecno. #1)}
\bigbreak\noindent{\it\secsym\the\subsecno. #1}\writetoca{\string\quad
{\secsym\the\subsecno.} {#1}}\par\nobreak\medskip\nobreak}
\def\appendix#1#2{\global\meqno=1\global\mthno=1\global\mexno=1%
\global\mquno=1
\global\subsecno=0
\xdef\secsym{\hbox{#1.}}
\bigbreak\bigskip\noindent{\bf Appendix #1. #2}\message{(#1. #2)}
\writetoca{Appendix {#1.} {#2}}\par\nobreak\medskip\nobreak}
%
%
\def\thm#1{\xdef #1{\secsym\the\mthno}\writedef{#1\leftbracket#1}%
\global\advance\mthno by1\wrlabeL#1}
\def\que#1{\xdef #1{\secsym\the\mquno}\writedef{#1\leftbracket#1}%
\global\advance\mquno by1\wrlabeL#1}
\def\exm#1{\xdef #1{\secsym\the\mexno}\writedef{#1\leftbracket#1}%
\global\advance\mexno by1\wrlabeL#1}
%
%
\def\ref{\the\refno\nref}
\def\nref#1{\xdef#1{\the\refno}\writedef{#1\leftbracket#1}%
\ifnum\refno=1\immediate\openout\rfile=refs.tmp\fi
\global\advance\refno by1\chardef\wfile=\rfile\immediate
\write\rfile{\noexpand\item{[#1]\ }\reflabeL{#1\hskip.31in}\pctsign}\findarg}
\def\bref{\nref}
%

%
%
%

%
%
\def\al{\alpha} \def\be{\beta}   
\def\de{\delta}  \def\De{\Delta} 
  \def\et{\eta} \def\th{\theta}
    
 \def\La{\Lambda} \def\rh{\rho} 
  \def\up{\upsilon}  
\def\ph{\phi}     
     
%
%

%
 
%
%
  \def\cF{{\cal F}} 
   
  \def\cW{{\cal W}} 
%
%
%
\def\lefthook{{\vrule height5pt width0.4pt depth0pt}}
\def\righthook{{\vrule height5pt width0.4pt depth0pt}}
\def\leftrighthookfill{$\mathsurround=0pt \mathord\lefthook
     \hrulefill\mathord\righthook$}
\def\underhook#1{\vtop{\ialign{##\crcr$\hfil\displaystyle{#1}\hfil$\crcr
      \noalign{\kern-1pt\nointerlineskip\vskip2pt}
      \leftrighthookfill\crcr}}}
%
%


\def\ie{{\it i.e.\ }}
\def\eg{{\it e.g.\ }}

\def\ZZ{Z\!\!\!Z}               
\def\NN{I\!\!N}                 %
             %




\hfuzz=10pt
\nopagenumbers
\pageno=0

\def\cF{F}
\def\ffM{ \hbox{$M$}\kern-.9em\hbox{$\overline{\phantom{N}}$}}
\def\hep#1{{\tt hep-th/{#1}}}

\def\PL{Phys.\ Lett.\ }
\def\NP{Nucl.\ Phys.\ }
\def\CMP{Comm.\ Math.\ Phys.\ }

\bref\LZs{
B.H.~Lian and G.J.~Zuckerman, \PL {\bf 254B} (1991) 417;
\PL {\bf 266B} (1991) 21; \CMP {\bf 145} (1992) 561.}

\bref\BMPa{
P.~Bouwknegt, J.~McCarthy and K.~Pilch, \CMP {\bf 145} (1992) 541.}

\bref\BLNWa{
M.~Bershadsky, W.~Lerche, D.~Nemeschansky and N.P.~Warner,
\PL {\bf 292B} (1992) 35.}

\bref\BLNWb{
M.~Bershadsky, W.~Lerche, D.~Nemeschansky and N.P.~Warner,
{\it Extended $N=2$ superconformal structure of gravity and
$W$-gravity coupled to matter}, CALT-68-1832, \hep{9211040}.}

\bref\BMPe{
P.~Bouwknegt, J.~McCarthy and K.~Pilch, {\it Semi-infinite
cohomology of $\cW$-algebras}, USC-93/11, ADP-93/200/M15,
\hep{9302086}.}

\bref\Pope{
H.~Lu, C.N.~Pope, S.~Schrans and X.J.~Wang, {\it On the spectrum and
scattering of $W_3$ strings}, CTP~TAMU-4/93, \hep{9301099}.}

\bref\West{
M.~Freeman and P.~West, {\it The covariant scattering
and cohomology of $\cW_3$ string}, KCL-TH-93-2, \hep{9302114}.}

\bref\Za{
A.B.~Zamolodchikov, Theor.\ Math.\ Phys.\ {\bf 65} (1985) 1205.}

\bref\BS{
P.~Bouwknegt and K.~Schoutens, Phys. Rep. {\bf 223} (1993) 183.}

\bref\FKW{
E.~Frenkel, V.G.~Kac and M.~Wakimoto, \CMP {\bf 147} (1992) 295.}

\bref\Math{
S.\ Wolfram,  {\it Mathematica: a system for doing mathematics by
computer}, Addison Wesley, 1988.}

\bref\TM{
J.~Thierry-Mieg, \PL {\bf 197B} (1987) 368.}

\bref\Fe{
B.L.~Feigin, Usp.\ Mat.\ Nauk {\bf 39} (1984) 195.}

\bref\FGZ{
I.B.~Frenkel, H.~Garland and G.J.~Zuckerman,
Proc.\ Natl.\ Acad.\ Sci. USA {\bf 83} (1986) 8442.}

\bref\BMPc{
P.~Bouwknegt, J.~McCarthy and K.~Pilch, {\it Semi-infinite cohomology
in conformal field theory and 2d gravity}, CERN-TH.6646/92,
\hep{9209034}.}

\bref\Wi{
E.~Witten, \NP {\bf B373} (1992) 187;
E.~Witten and B.~Zwiebach, \NP {\bf B377} (1992) 55.}

\bref\BMPb{
P.~Bouwknegt, J.~McCarthy and K.~Pilch, {\it Some aspects of free field
resolutions in $2D$ CFT with application to the quantum Drinfeld-Sokolov
reduction}, in proc. ``Strings and Symmetries 1991,'' eds. N.~Berkovitz
et.\ al., World Scientific, 1991.}

%
%

\line{}
\vskip1cm
\def\title{On the BRST structure of $\cW_3$ gravity coupled to
$c=2$ matter}

\centerline{{\bf  \title }\footnote{$^\flat$}{To appear in the
proceedings
of the AMS Special Session on
``Geometry and Physics,'' USC, Los Angeles,
November 5-6, 1992.} }
\vskip1cm

\centerline{Peter Bouwknegt$^1$,
Jim McCarthy$^2$
and Krzysztof Pilch$^1$}
\bigskip

\centerline{\sl $^1$ Department of Physics and Astronomy }
\centerline{\sl  University of Southern California}
\centerline{\sl Los Angeles, CA~90089-0484, USA}
\medskip

\centerline{\sl $^2$ Department of Physics and Mathematical Physics}
\centerline{\sl University of Adelaide, Adelaide, SA~5001, Australia}
\vskip1.5cm

\def\abstr{We present some explicit results on the structure of
singular vectors in $c=2$ Verma modules of the $\cW_3$ algebra. Using the
embedding patterns of those vectors we construct resolutions for the
$c=2$ irreducible modules, and thus are able to compute some of the BRST
cohomology of $\cW_3$ gravity coupled to $c=2$ matter.  In particular, we
determine the states in the ground ring of the theory.}

\centerline{\bf Abstract}\smallskip
\abstr
\noindent

 \vskip1.5cm

\vfil
\line{USC-93/14 \hfil}
\line{ADP-93-203/M17 \hfil}
\line{{{\tt hep-th/9303164}}\hfil March 1993}

\eject

\footline{\hss\tenrm\folio\hss}

\font\ninerm=cmr9

\baselineskip=12pt
\centerline{{ \bf \title }}
\vskip.8truecm
\centerline{{\ninerm Peter  BOUWKNEGT$^1$, Jim  McCARTHY$^2$ and
Krzysztof  PILCH$^1$}}
\smallskip
\centerline{{\it  $^1$ University of Southern California,
Los Angeles, CA 90089, USA}}
\centerline{{\it  $^2$ University of Adelaide,
Adelaide, SA 5001, Australia}}

\bigskip
\vbox{\hbox{\centerline{{\ninerm ABSTRACT}}}
{\smallskip\leftskip 3pc \rightskip 3pc \noindent \ninerm \abstr\smallskip}}

\footline={\hss\tenrm\folio\hss}

\baselineskip=1.2 \baselineskip

\def\Wt{{\cal W}_3} \def\Wtp{{\cal W}_{3}^{(+)}}
\def\Wtz{{\cal W}_{3,0}}   \def\Wtm{{\cal W}_{3}^{(-)}}

\bigskip

\newsec{Introduction}

The BRST quantization of two dimensional gravity coupled to conformal matter
has posed the interesting mathematical problem of computing the semi-infinite
cohomology of the Virasoro algebra with values in a tensor product of
two highest weight modules [\LZs,\BMPa].
In this note we report on
some explicit results in a partial solution to the similar problem
of quantizing $\Wt$ gravity coupled to $\Wt$ matter.

More precisely, we study the cohomology of the BRST operator
recently constructed in [\BLNWa], on the product of two two-scalar Fock
space modules of the $\Wt$ algebra. This so called `$d=(2,2)$
$\Wt$ string' is a natural generalization of the usual $d=2$ string.

The main source of differences between the present problem and its analogue
for the Virasoro algebra is the nonlinear nature of the $\Wt$ algebra.
As a result, many of the  computational techniques of Lie algebra
cohomology either cannot be used, or are too difficult to implement due
to the much greater algebraic complexity of the problem.
In addition,  the structure of  positive energy modules of the
$\Wt$ algebra,  most notably of $c=2$ Verma modules,  has only
partially been understood.

This note is organized as follows. In Section 2, apart from basic
definitions,  we introduce positive energy modules of the $\Wt$ algebra,
and summarize some of the known results. In Section 3 we discuss in some
detail the  structure of Verma modules of the $\Wt$ algebra  at $c=2$. In
particular, we summarize our explicit studies of singular vectors
at low-lying weights in those modules, and give
their embedding patterns. From those
results we conjecture resolutions for a class of irreducible modules
in terms of (generalized) Verma modules. Finally, in Section 4 we introduce
the BRST complex of the $\Wt$ algebra and discuss its cohomology on the
tensor product of two two-scalar Fock spaces. For other and/or related
results we refer the reader to \eg [\BLNWb-\West] and
references therein.

\newsec{The $\Wt$ algebra and its modules}

The $\Wt$ algebra is generated by the
operators $L_m$, $W_m$, $ m\in\ZZ$,  which
satisfy the following commutation relations [\Za]
(for a review on $\cW$-algebras see [\BS] and references therein).%
\foot{We follow here the physicists
convention in which the central element, $c$,
of the algebra is set to a constant.}
\eqnn\aavir \eqnn\aavw \eqnn\aaww
$$\eqalignno{
[L_m,L_n]&=(m-n)L_{m+n}+\textstyle{c\over 12}
m(m^2-1)\de_{m+n,0}\,,&\aavir\cr
[L_m,W_n]&= (2m-n) W_{m+n}  \,, &\aavw\cr
[W_m,W_n]&=  (m-n)\left( \textstyle{1\over 15}
(m+n+3)(m+n+2)-\textstyle{1\over 6}
(m+2)(n+2)\right)L_{m+n} &\cr
& +\be (m-n) \La_{m+n} +\textstyle{c\over 360}m(m^2-1)(m^2-4)\de_{m+n,0}
	     \,, &\aaww\cr}$$
where $\be=16/(22+ 5c)$ and
\eqn\aalam{
\La_{m}=\sum_n (L_{m-n} L_n) -\textstyle{3\over 10}(m+3)(m+2) L_m\,,}
with the normal ordered product given by
\eqn\aanop{
(L_mL_n)=\cases{L_mL_n& for $m\leq -2$,\cr
		L_nL_m& for $m> -2$.\cr}}

As for Lie algebras, the commutators  satisfy the Jacobi identities, but
the normal ordered terms introduce a nonlinear structure into the algebra.
The $\Wt$ algebra clearly contains the
Virasoro algebra, generated by the $L_n$, $n \in \ZZ$,  as a Lie subalgebra.
The maximal abelian subalgebra (`Cartan subalgebra')
$\Wtz$ of $\Wt$ is spanned by  $ L_0$ and $ W_0 $, but,
 because   $({\rm ad}\,W_0)$
is not diagonalizable,  $\Wt$ does not admit a root space decomposition.
Nevertheless, it is still convenient to decompose the generators
of $\Wt$ according to the  $(-{\rm ad}\,L_0)$ eigenvalue, and, in particular,
define
$\Wt^{(\pm)}=\{L_n,W_n\,|\, \pm n>0\}$.

In the following we will consider only the so-called positive energy
modules of $\Wt$ [\FKW],
which  are defined by the condition that (the energy operator) $-L_0$
is diagonalizable
with finite dimensional eigenspaces, and with the spectrum bounded from below.
If the lowest energy eigenspace is one dimensional we
denote the eigenvalue  of $L_0$ and $W_0$
on this highest weight state by $h$ and $w$, respectively.

In particular, the generalized Verma module $M(h,w,c)_N$ is defined
as the positive energy module induced from an $N$-dimensional
indecomposable representation of $\Wtz$. More explicitly, let
$v_0,\ldots,v_{N-1}$ be a canonical basis such that \eqnn\aagha \eqnn\aaghb
$$\eqalignno{
L_0 v_i&=hv_i\,,\quad  i=0,\ldots,N-1\,,& \aagha\cr
W_0v_0 &= wv_0\,,\quad W_0 v_i =wv_i+v_{i-1}\,,\quad
i=1,\ldots,N-1\,.&\aaghb\cr}$$
Then $M(h,w,c)_N$ is spanned by the monomials of the form
\eqn\aaverb{L_{-n_1}\ldots L_{-n_i}W_{-m_1}\ldots W_{-m_j} v_k\,,\quad
i,j\geq 0 \,,\quad k=0,\ldots,N-1\,,}
on which the generators in $\Wtm$ act freely, while the action of those in
$\Wtp$ and $\Wtz$ is  determined using \aavir-\aaww, \aagha-\aaghb\ and
\eqn\aavan{L_nv_i=W_nv_i=0\,,\quad n>0\,,\quad i=0,\ldots,N-1\,.}

Clearly, for $N=1$ we  recover the usual definition of the Verma module
$M(h,w,c)$. By the standard argument,  $M(h,w,c)$ contains a maximal
submodule. We denote the corresponding irreducible quotient
module  by $L(h,w,c)$.
The module contragradient to $M(h,w,c)$ will be denoted by
$\ffM(h,w,c)$.

{}From the physics viewpoint the most natural class of
positive energy
modules of $\Wt$ are the Fock space modules $F(\La,\al_0)$, which
arise in the free field realization of the $\Wt$ algebra in terms of two scalar
fields,\foot{For a detailed discussion  see [\BS] and references therein.}
and thus generalize the Feigin-Fuchs modules of the Virasoro algebra.
Here $\La$ is an $sl(3)$ weight. [In the following we will
use the notation $P_+$ for the set of dominant integral weights of
$sl(3)$, $\De_+$ for its positive roots, and $Q_+=\ZZ_+\De_+$ for
the positive root lattice.] The background charge, $\al_0 $, is
determined by $c=2-24\al_0^2$. The  eigenvalues of $L_0$ and $W_0$ on the
highest weight state $|\La,\al_0\rangle$ of $F(\La,\al_0)$ are
\eqn\aawghts{
\eqalign{
h(\La)&=-(\th_1\th_2+\th_1\th_3+\th_2\th_3)-\al_0^2=\half(\La,\La+
2\al_0\rh)\,,\cr
w(\La)&=\sqrt{3\be}\,\th_1\th_2\th_3\,,\cr
 }}
where
\eqn\aalamb{\th_1=(\La+\al_0\rh,\La_1)\,,\quad
	    \th_2=(\La+\al_0\rh,\La_2-\La_1)\,,\quad
	    \th_3=(\La+\al_0\rh,-\La_2)\,.}
The weights $\La_1$ and $\La_2$ are the fundamental weights of $sl(3)$, and
$\rh=\half\sum_{\al\in\De_+}\al$ is the Weyl vector.
Note that $h(\La)$ and $w(\La)$ as in  \aawghts\  determine $\La$ only up to a
Weyl rotation $\La\rightarrow w(\La+\al_0\rh)-\al_0\rh$, $w\in W$.
In the following we will also write $M(\La,c)$, or simply $M(\La)$ for
the (generalized) Verma module $M(h(\La),w(\La),c)$.

The following two theorems summarize some of the known results on the structure
of Fock space modules $F(\La,\al_0)$. We have written them in a form
which allows an immediate generalization to higher rank $\cW$ algebras.
The first theorem can be  proven by examining the Shapovalov form on  the
Verma module [\BMPe],  while the second one essentially follows from the
unitarizability of $F(\La,0)$, an explicit construction of singular vectors,
and direct comparison of the character for each side [\BMPe].
\thm\thBb
\proclaim Theorem \thBb. {
\item{(a)}
$
\cF(\La,\al_0) \ \cong\ \cases{
M(\La,c) & if  $(\La+\al_0\rh,\al)\notin
(\NN\al_+ + \NN\al_-)$ for all $\al\in\De_+\,,$\cr
{\ffM}(\La,c) & if $(\La+\al_0\rh,\al)\notin
-(\NN\al_+ + \NN\al_-)$ for all $\al\in\De_+\,.$ \cr}$
\smallskip
\item{}
In particular, if $(\La+\al_0\rh,\al) \notin (\NN\al_+ + \NN\al_-)$
for all $\al\in\De$, then $M(\La,c)$ (and thus also $\cF(\La,\al_0)$)
is irreducible.
\item{(b)}
For $\al_0{}^2 \leq-4$ or, equivalently,
$c\geq c_{crit}-\ell =  \ell + 48 |\rh|^2$ we have
$$
\cF(\La,\al_0) \cong \cases{
M(\La,c) & for $i(\La+\al_0\rh) \in \et D_+\,,$ \cr
{\ffM}(\La,c) & for $-i(\La+\al_0\rh) \in \et D_+\,,$ \cr}
$$
where $\ell $ is the rank of the $\cW$ algebra
($\ell=2$ for $\Wt$),  $D_+$ denotes
the fundamental Weyl chamber, and $\et = {\rm sign}(-i\al_0)$.}

\thm\thBc
\proclaim Theorem \thBc. If $w\in W$ such that $w\La\in P_+$ then
\eqn\eqBd{
\cF (\La,0) = \bigoplus_{\scriptstyle \be\in Q_+\atop \scriptstyle
w\La+\be\in P_+}
m(w\La;\be)\,L(w\La+\be)\,,
}
where, for  $\La\in P_+$ and $\be\in Q_+$, the multiplicity $m(\La;\be)$
(with which the $c=\ell=2$ irreducible $\Wt$-module $L(\La+\be)$
occurs in the direct sum decomposition of $\cF(\La,0)$) is
equal to the multiplicity of the weight $\La$ in the irreducible
finite dimensional representation of $sl(3)$ with highest weight
$\La+\be$.

\newsec{ The structure of $\Wt$ Verma modules for $c=2$ }

It is an interesting mathematical problem, as well as being crucial for
our results in the next section, to construct resolutions of the
irreducible  modules $L(\La,c)$ in terms of Verma modules.
For that reason we have studied explicitly the
singular vectors and their embedding structure
in $c=2$ (generalized) Verma modules $M(\La)_N$
for low lying weights $\La$. Except for the simplest cases, all the
computations were performed using algebraic manipulation
programs [\Math].
We discuss below our results, which are summarized in
Tables I, Ia, II and IIa in Appendix A.

For convenience we denote weights by their Dynkin labels $(a_1, a_2)$,
instead of $\La$, where $\La=a_1\La_1+a_2\La_2$. We recall that, as  usual,
singular vectors are defined to be those vectors in $M(\La)$ that are
annihilated by all the generators
in $\Wtp$. The vectors in the tables are such that $L_0$ is diagonal and $W_0$
is in a canonical Jordan form. Then $u_{a_1a_2}$ denotes a singular vector at
energy level $h(a_1,a_2)$, which is an eigenstate of $W_0$ with the eigenvalue
$w(a_1,a_2)$. Similarly, $v_{a_1a_2}$ is a singular vector for which
$W_0v_{a_1 a_2}=w(a_1,a_2) v_{a_1 a_2} + u_{a_1 a_2}$. The $v$-type singular
vectors are determined only up to addition of an $u$-type vector. [Although
one might expect to also find higher dimensional indecomposable representations
of $\Wtz$ on the subspaces of singular vectors, it is a curious fact that
they do not arise in the examples
below. However, in  many cases a singular vector $u$ or a doublet
$(v,u)$ is found inside a higher dimensional indecomposable representation
of $\Wtz$.]

A singular vector $u_{a_1a_2}$ in $M(b_1, b_2)_N$ determines a canonical
embedding of $M(a_1,a_2)$ into $M(b_1, b_2)_N$. Similarly, any $v_{a_1 a_2}$
in $M(b_1, b_2)_N$ defines a homomorphism of $M(a_1,a_2)_2$ into $M(b_1,
b_2)_N$
which, in general, has a nontrivial kernel. Singular vectors which
vanish under such homomorphism are denoted by primes; \eg in
Table I,   $u_{30}'$ and $u_{03}'$ in $M(1,1)_2$ vanish in the submodule
generated by $(v_{11},u_{11})$ in $M(0,0)$.  By examining
various columns for the same entries one can also determine singular vectors
that lie in intersections of submodules, \eg from
Tables II and IIa we find that the
intersection of $M(2,0)$ and $M(1,2)_2$ in $M(0,1)$ contains $u_{12}$ and
$(v_{31},u_{31})$.

The singular vectors and their embedding patterns lead to
the following conjecture for resolutions of irreducible
modules $L(a_1,a_2)$ at $c=2$.

\thm\resconj
\proclaim Conjecture \resconj. The resolution of an
irreducible module $L(\La,2)$ in terms of generalized Verma modules
is one of three types as given in Appendix B. The type of
the resolution depends on whether $\La$ lies in the interior of the
fundamental Weyl chamber, on the boundary, or, for $\La=0$, on the intersection
of two boundaries.

Note that these resolutions are finite, as is also the case
for resolutions of $c=2$ irreducible modules in
terms of Fock space modules.
It is a nontrivial check that the resolutions in Appendix B
give rise to characters which agree with those obtained from the Fock
space resolutions.
Our explicit results confirm those resolutions
for the weights $(0,0)$, $(0,1)$, $(1,0)$,
$(1,1)$, $(0,2)$ and $(2,0)$, and partially for some of the  other
weights.  However, it is possible, though unlikely,
that the resolutions may get modified
because of unaccounted singular vector
structure beyond the levels which we have studied explicitly.

\newsec{The BRST cohomology}

\def\ghbo{b^{[1]}} \def\ghco{c^{[1]}}
\def\ghbt{b^{[2]}} \def\ghct{c^{[2]}}

The basic idea of the explicit  construction of the semi-infinite cohomology
complex of the $\Wt$-algebra [\TM,\BLNWa]
(in the physics literature called the
BRST complex) is essentially the same as for the Virasoro
or affine Lie algebras (see \eg [\Fe-\BMPc]).  First one introduces an infinite
dimensional Clifford algebra of ghost operators, $(\ghbo_m,\ghco_m)$
and  $(\ghbt_m,\ghct_m)$, corresponding to the generators $L_m$ and $W_m$,
$m\in\ZZ$, respectively. These ghost operators satisfy the
anti-commutation relations $\{\ghco_m ,\ghbo_n\}=\de_{m+n,0}$ and
$\{\ghct_m ,\ghbt_n\}=\de_{m+n,0}$. The corresponding standard positive energy
module, $F^{gh}$,
of the ghost algebra
is freely generated by the negative mode operators acting on the  highest
weight state $|0\rangle_{gh}$ satisfying
\eqn\ghvac{
\eqalign{\ghbo_n |0\rangle_{gh}&=
\ghbt_n |0\rangle_{gh}=0\,,\quad n\geq 0\,,\cr
\ghco_n |0\rangle_{gh}&=
\ghct_n |0\rangle_{gh}=0\,,\quad n>0\,.\cr}}
\def\rgh{{\rm gh}}
This  ghost Fock space  has a canonical grading by the ghost number
${\rm gh}(\cdot)$, where
$\rgh(\ghco_n)=\rgh(\ghct_n)=-\rgh(\ghbo_n)=-\rgh(\ghbt_n)=1$,  while the ghost
number of the vacuum state $|0\rangle_{gh}$ is zero. Given a vector space
$V$ on which the $\Wt$ algebra acts, the space of semi-infinite forms is simply
the tensor product $V\otimes F^{gh}$, with the degree given by the ghost
number. The problem is then to  construct a nilpotent operator of degree one,
the BRST operator, which becomes the differential in the complex. In addition,
the structure of the leading terms should be the same as for the
ordinary differential in the semi-infinite complex of Lie algebras.
The two cases in which this construction has been carried
out, and which appear to be relevant in  $\Wt$ gravity theories, are:
\smallskip
\item{i)} $V$ is a $\Wt$ algebra module with the central charge $c=100$ [\TM].
\item{ii)} $V$ is a tensor product of {\it two} $\Wt$-modules,
$V=V^M\otimes V^L$, with  $c^M+c^L=100$ [\BLNWa].
\smallskip
\noindent
In the second case, the differential is given by the following
normal ordered operator
\eqn\bbrst{ \eqalign{
d
&=\sum_{m}\big( \ghct_{-m}(\widetilde W^M_m - i \widetilde W^L_m) +
		\ghco_{-m}(L^M_m+L^L_m) \big)\cr
&+\sum_{m,n}\big(
-\half(m-n)\,\ghco_{-m}\ghco_{-n}\ghbo_{m+n}
-(2m-n)\ghco_{-m}\ghct_{-n}\ghbt_{m+n}\cr
&\quad\quad\quad\,\,
-\textstyle{1\over 6}\mu(m-n)(2m^2-mn+2n^2-8)\,
\ghct_{-m}\ghct_{-n}\ghbo_{m+n}\big)\cr
&+\sum_{m,n,p}\big(-\half(m-n)\, \ghct_{-m}\ghct_{-n}\ghbo_{-p}
(L^M_{m+n+p}-L^L_{m+n+p})\big)\,,\cr}
}
where $\widetilde W^{M,L}=W^{M,L}/\sqrt{\be^{M,L}}$ and
$\mu=(1-17\be^M)/(10\be^M)$.
We will denote the cohomology of $d$ at  degree $n$ by
$H^{(n)}(d,V^M\otimes V^L)$, and call it the semi-infinite
cohomology, or the BRST cohomology, of the $\Wt$  algebra on $V^M\otimes V^L$.
When either $V^M$ or $V^L$ is a trivial module, the BRST operator \bbrst\
reduces to the  one of [\TM].

In  the context of $\Wt$ gravity  one is interested in computing
the BRST cohomologies when $V^M$ is either an  irreducible
module $L(h^M,c^M)$ or a Fock space module $F(\La^M,\al_0^M)$. In both
cases $V^L$ is a Fock space module $F(\La^L,\al_0^L)$.  The cohomology
of $L(h^M,c^M)\otimes F(\La^L,\al_0^L)$, when $c^M<2$, for the so-called
$\Wt$ minimal models coupled to $\Wt$ gravity, has been discussed in [\BMPe].
In the following we will consider only the second case when $c^M=2$,
\ie the cohomology of $F(\La^M,0)\otimes F(\La^L,2i)$.

There are several reasons why understanding this cohomology appears to
be important. Firstly, it  gives the spectrum of physical states of the
$d=(2,2)$ $\Wt$ string, which one expects to have a similar structure
to that of the $d=2$ string. In particular, the nontrivial cohomology
should arise  only in degrees between $-3$ and $5$, and the states with
the lowest, \ie $-3$, ghost number form a ``ground ring'' which to a
large extent reflects the properties of the theory [\Wi]. The other
reason is that, provided our intuition from the case of the Virasoro
algebra is correct, one may be able to continuously deform the
cohomology of two Fock space modules to other  values of $c^M$.
[In fact the explicit construction of some cohomology states in
[\BLNWb] strongly supports such a possibility.] If this is indeed true,
then all the interesting cohomologies could easily be computed because
the  resolutions of  irreducible modules in terms of Fock space modules
are known [\BMPb,\FKW].

A direct computation of the cohomology of two Fock space modules
appears to be quite difficult (unlike the Virasoro algebra [\BMPa]) because
of the algebraic complexity of the problem. However, given the
results of Sections 2 and 3,  we may obtain partial results by reducing
the computation to the one of Verma modules. In particular,
a direct modification of standard arguments  yields the
the following `reduction theorem'  [\BMPe]
\thm\redthm
\proclaim Theorem \redthm. The cohomology $H(d,
M(\La^M)_N\otimes \ffM(\La^L))$ is nonvanishing
if and only if  $w(\La^M+\al_0^M\rh)=-i(\La^L+\al_0^L\rh)$ for some
$w\in W$, in which case it is spanned by  the states $v_0$, $\ghco_0 v_0$,
$\ghct_0 v_{N-1}$ and $\ghco_0 \ghct_0 v_{N-1}$,
where $v_i=v_i^M\otimes \overline v^L\otimes |0\rangle_{gh}$  ({\it cf}\
\aagha\ -- \aavan).

 Let us  define $\al_\pm=\half(\al_0^M\mp i \al_0^L)$ so
that $\al_+\al_-=-1$. An immediate consequence of Theorems \thBb\ and
\redthm\ is that the cohomology is essentially trivial for a
generic class of Fock spaces.
More precisely, we parametrize the momenta $\La^M$ and $\La^L$  by
\eqn\DDbb{\eqalign{
\La^M+\al_0^M\rh &= \al_+\La^{(+)} + \al_-\La^{(-)}\,,\cr
-i(\La^L+\al_0^L\rh) &= \al_+\La^{(+)} - \al_-\La^{(-)}\,.\cr}}
We call them  generic iff there is no positive root
$\al\in \De_+$ such that
\eqn\DDbc{(\La^{(+)},\al)\in\ZZ\,,\quad (\La^{(-)},\al)\in\ZZ\,,\qquad
(\La^{(+)},\al) (\La^{(-)},\al)>0\,.}
Then we have
\thm\thDDa
\proclaim Theorem \thDDa. For generic momenta   $\La^M$ and $\La^L$
as defined above,
$H(d,\cF(\La^M,\al_0^M)\otimes \cF(\La^L,\al_0^L))
\not=0$ iff there exists $w\in W$ such that $w(\La^M+\al_0^M\rh)=
-i (\La^L+\al_0^L\rh)$, in which case it is spanned by the states
$v$, $c_0^{[1]}v$,
$c_0^{[2]}v$ and
$c_0^{[1]}c_0^{[2]}v$,
where $v=|\La^M\,\rangle\otimes |\La^L\,\rangle
\otimes |0\,\rangle_{gh}$ is the physical vacuum.

Returning to the $c^M=2$ case (\ie $\al_\pm=\pm1$) we first note that
the structure of the cohomology simplifies
considerably because of the following observation
\thm\thDa
\proclaim Theorem \thDa. The cohomology spaces $H^{(n)}(d,\cF(\La^M,0)
\otimes \cF(\La^L,2i))$ carry a fully reducible representation of
$sl(3)$. The $sl(3)$ generators are explicitly given by the zero modes
of a level-1
Frenkel-Kac-Segal vertex operator construction in terms of matter
fields only.

We now consider a non-generic case, when both $\La^{M}$ and
$-i\La^{L}$ are integral weights.\foot{The remaining cases, when
$\La^{M,L}$ are integral along certain roots, can be analyzed
similarly.}
Moreover, we further restrict the Liouville momenta to the
region $-i\La^L+2\rh \in  D_+$. Then
we can combine Theorems \thBb\ and \thBc\ and reduce the computation of
$H(d,\cF(\La^M,0)\otimes\cF(\La^L,\al_0^L))$ to that of
$H(d,L(\La)\otimes {\ffM}(\La^L))$
for the $c=2$ irreducible modules,
$L(\La)$, which appear in the decomposition of $\cF(\La^M,0)$.  In turn,
the latter cohomology  can be determined  by  standard
arguments using the resolutions of Section 3 and Theorem \redthm \ -- in
particular one finds that indeed the lowest ghost number states  have
ghost number $-3$. From this follows directly the
enumeration of all the states in the corresponding (chiral) ground
ring $\cal R$, which we summarize as \thm\thDb
\proclaim Conjecture \thDb.
Given a pair of integral weights $(\La^M,-i\La^L)$,
let $w\in W$ be such that $w\La^M\in P_+$. Then there are
ground ring elements $\ph_{(\La^M,-i\La^L)}$ provided
$w\La^M + \be =-i\La^L$ for some $\be\in Q_+$. The multiplicity  depends
on whether $w\La^M+\be$ is in the interior or on the boundary of the
fundamental Weyl chamber, and is given by
$2m(w\La^M;\be)$ or $m(w\La^M;\be)$, respectively.
The (relative) energy level at which $\ph_{(\La^M,-i\La^L)}$
 occurs is given by
$\half|-i\La^L + 2\rh|^2 - \half |\La^M|^2$.

It is not difficult to see that under the $sl(3)$ symmetry in
Theorem \thDa\ the ground ring  $\cal R$
decomposes into the following direct
sum of irreducible finite-dimensional  $sl(3)$ modules
\eqn\ringdec{
{\cal R}= \bigoplus_{\La\in P_+}  {\cal R}^{(1)}_\La
\oplus \bigoplus_{\La\in P_{++}} {\cal R}^{(2)}_\La\,,}
\ie each representation corresponding to an integral  weight
in the interior of the fundamental Weyl chamber ($P_{++}$)
occurs exactly twice, while those on the boundary
($P_{+}\setminus P_{++}$) exactly once.
Since the $sl(3)$ symmetry acts  by automorphisms of the
ground ring, a simple counting of the states indicates
that the generators of the ring will have to satisfy vanishing
relations.

\newsec{Acknowledgements}

P.B.\ acknowledges the support of the Packard Foundation,
J.M.\ that
of the Australian Research Council, while K.P.\ is supported in
part by the Department of Energy Contract \#DE-FG03-84ER-40168.
We would like to thank L.\ Romans for supplying routines for
working with conformal fields at the level of modes.

\footatend\immediate\closeout\rfile\writestoppt
\baselineskip=14pt{\bigskip\noindent {\bf  References}}%
\bigskip{\frenchspacing%
\parindent=20pt\escapechar=` \input refs.tmp\vfill\eject}\nonfrenchspacing

\vfill\eject
\noindent
{\bf Appendix A}
\bigskip
\noindent
{Table I }
\vskip .5cm
\begintable
$\,\,h\quad$| $M(0,0)$ | $M(1,1)$ | $M(3,0)$ | $M(0,3)$ |
$M(2,2)$ | $M(4,1)$ | $M(1,4)$ | $M(3,3)$ | $\cdots$ \cr
0 | $u_{00}$   ||||||||\nr
1 | $(v_{11},u_{11})$ | $u_{11}$ |||||||\nr
3 | $u_{30}$ | $u_{30}$ | $u_{30}$ | |||||\nr
  | $u_{03}$ | $u_{03}$ | | $u_{03}$ | | | | | \nr
4 | $(v_{22}, u_{22})$ | $(v_{22}, u_{22})$ | $u_{22}$ | $u_{22}$
  | $u_{22}$ ||||\nr
7 | $\vdots$ |  $(v_{41}, u_{41}) $ |
    $(v_{41}, u_{41}) $  |  $ u_{41} $ | $u_{41}$
  | $u_{41}$ ||| \nr
  | $\vdots$ | $(v_{14} ,u_{14})$ | $u_{14}$ | $(v_{14} ,u_{14})$
  | $u_{14}$ || $u_{14}$ ||\nr
9 | $\vdots$ | $(v_{33}, u_{33})$ | $(v_{33}, u_{33})$ |
    $(v_{33}, u_{33})$ | $(v_{33}, u_{33})$ | $u_{33}$ | $u_{33}$ | $u_{33}$
|\nr
  $\vdots$ | $\vdots$ |$\vdots$ |$\vdots$ |$\vdots$ |$\vdots$ |$\vdots$
  |$\vdots$ |$\vdots$ |$\vdots$
\endtable

\bigskip
\bigskip
\noindent
{Table  Ia}
\vskip .5cm
\begintable
$\,\,h\quad$|   $M(1,1)_2$ | $M(2,2)_2$ | $M(4,1)_2$ | $M(1,4)_2$ |
	       $M(3,3)_2$ |  $\cdots$ \cr
1 | $(v_{11},u_{11})$ ||||| \nr
3 | $u_{30},\,u_{30}'$ |  ||||\nr
  | $u_{03},\,u_{03}'$ |  | | | | \nr
4 | $(v_{22},u_{22}),\,u'_{22}$ | $(v_{22},u_{22})$ ||||\nr
7 | $\vdots$ |  $(v_{41},u_{41})$
  | $(v_{41},u_{41})$ ||| \nr
  | $\vdots$ |  $(v_{14},u_{14})$ || $(v_{14},u_{14})$ ||\nr
9 | $\vdots$ | $(v_{33},u_{33}),\,u_{33}'$ | $(v_{33},u_{33})$ |
    $(v_{33},u_{33})$ | $(v_{33},u_{33})$ |\nr
     $\vdots$ |$\vdots$ |$\vdots$ |$\vdots$
  |$\vdots$ |$\vdots$ |$\vdots$
\endtable
\def\txt#1{$\hbox{$#1$}$}
\vfill\eject

\noindent
{Table  II}
\vskip .5cm
\begintable
$\,\,h\quad$| $M(0,1)$|
    $M(2,0)$ | $M(1,2)$ | $M(3,1)$ | $M(0,4)$ | $M(2,3)$ | $M(5,0)$
	     | $M(4,2)$ | $\cdots$ \cr
\txt{1\over 3} | $u_{01}$ ||||||||\nr
\txt{4\over 3} | $u_{20}$ | $u_{20}$ |||||||\nr
\txt{7\over 3} | $(v_{12},u_{12})$ | $u_{12}$ | $u_{12}$ ||||||\nr
\txt{13\over 3}| $(v_{31},u_{31})$ | $(v_{31},u_{31})$ | $u_{31}$ |
$u_{31}$ ||||| \nr
\txt{16\over 3}| $u_{04}$ |$u_{04}$ | $u_{04}$ |  | $u_{04}$ ||||\nr
\txt{19\over 3}| $(v_{23},u_{23})$ | $(v_{23},u_{23})$ |
$(v_{23},u_{23})$ | $u_{23}$
	       | $u_{23}$ | $u_{23}$ ||| \nr
\txt{25\over 3}| $\vdots$ |$u_{50}$ | $u_{50}$ |  $u_{50}$ | | | $u_{50}$ ||\nr
\txt{28\over 3}| $\vdots$ | $(v_{42},u_{42})$ | $(v_{42},u_{42})$ |
$(v_{42},u_{42})$ |
		 $u_{42}$ |  $ u_{42}$ | $ u_{42}$ |$ u_{42}$ |\nr
$\vdots$ |$\vdots$ |$\vdots$ |$\vdots$ | $\vdots$ |$\vdots$ |$\vdots$
|$\vdots$ |$\vdots$|$\vdots$
\endtable

\bigskip\bigskip

\noindent
{Table  IIa}
\vskip .5cm
\begintable
$\,\,h\quad$| $M(1,2)_2$ |
  $M(3,1)_2$ | $M(2,3)_2$ | $M(4,2)_2$ | $\cdots$ \cr
\txt{7\over 3} | $(v_{12},u_{12})$ ||||\nr
\txt{13\over 3}| $(v_{31},u_{31})$ | $(v_{31},u_{31})$ ||| \nr
\txt{16\over 3}| $u_{04},\,u_{04}'$ ||||\nr
\txt{19\over 3}| $(v_{23},u_{23}),\,u_{23}'$ |
		 $(v_{23},u_{23})$ | $(v_{23},u_{23})$  || \nr
\txt{25\over 3}| $\vdots$ |$u_{50},\,u_{50}'$ ||| \nr
\txt{28\over 3}| $\vdots$ |$(v_{42},u_{42}),\,u_{42}'$ | $(v_{42},u_{42})$ |
		 $(v_{42},u_{42})$ | \nr
$\vdots$ |$\vdots$ |$\vdots$ |$\vdots$ |$\vdots$ | $\vdots$
\endtable

\vfill\eject
\message{*** Please, remember about Appendix B! *** }
\end



\font\ttbf=cmbx10 scaled\magstep1

\def\up{\uparrow}
\pageno=12

\noindent
{\ttbf Appendix B}
\bigskip

\noindent
Resolution of  $ L(0,0)$:
$$\matrix{
0\cr\cr
\up\cr\cr
M(0,0)\cr\cr
\up\cr\cr
M(1,1)_2\cr\cr
\up\cr\cr
M(3,0)\oplus M(0,3)\cr\cr
\up\cr\cr
M(2,2)\cr\cr
\up\cr\cr
0\cr}
$$
\bigskip
\noindent
Resolutions of $  L(a_1,0)$ and
$ L(0,a_2)$, $  a_1,a_2>0$:
$$
\matrix{
0\cr\cr
\up\cr\cr
M(a_1,0)\cr\cr
\up\cr\cr
M(a_1+1,1)_2\oplus M(a_1-1,2)\cr\cr
\up\cr\cr
M(a_1+3,0)\oplus M(a_1+1,1)\oplus M(a_1,3)_2\cr\cr
\up\cr\cr
M(a_1+2,2)_2\oplus M(a_1,3)\cr\cr
\up\cr\cr
M(a_1+2,2)\cr\cr
\up\cr\cr
0\cr}\,  \quad\quad\quad
\matrix{
0\cr\cr
\up\cr\cr
M(0,a_2)\cr\cr
\up\cr\cr
M(2,a_2-1)\oplus M(1,a_2+1)_2\cr\cr
\up\cr\cr
M(3,a_2)_2\oplus M(1,a_2+1)\oplus M(0,a_2+3)\cr\cr
\up\cr\cr
M(3,a_2)\oplus M(2,a_2+2)_2\cr\cr
\up\cr\cr
M(2,a_2+2)\cr\cr
\up\cr\cr
0\cr}
$$
\vfill\eject

\noindent
Resolutions of $L(a_1,a_2)$, $a_1,a_2>0$:
\smallskip
$$
\matrix{
0\cr\cr
\up\cr\cr
M(a_1,a_2)\cr\cr
    \up    \cr\cr
  M(a_1+2,a_2-1) \oplus M(a_1+1,a_2+1)_2 \oplus M(a_1-1,a_2+2)  \cr\cr
    \up    \cr\cr
M(a_1+3,a_2)_2 \oplus M(a_1+1,a_2+1) \oplus M(a_1+2,a_2+2)
 \oplus M(a_1+1,a_2+1) \oplus M(a_1,a_2+3)_2\cr\cr
    \up    \cr\cr
 M(a_1+3,a_2) \oplus M(a_1+2,a_2+2)_2 \oplus M(a_1+2,a_2+2)_2
\oplus M(a_1,a_2+3) \cr\cr
    \up    \cr\cr
   M(a_1+2,a_2+2) \oplus M(a_1+2,a_2+2)   \cr \cr\up \cr\cr0\cr}$$

\bigskip

\vfill\eject\end